\documentclass[12pt,preprint]{aastex}
\usepackage{graphics}

\begin{document}
\slugcomment{Submitted to ApJL}

\title{On the Nature of the Progenitor of the Type Ia SN2011fe in M101}

\author{Jifeng Liu\altaffilmark{1,2,3}, Rosanne Di Stefano\altaffilmark{2}, 
Tao Wang\altaffilmark{4,2}, and Maxwell Moe\altaffilmark{2}}
\altaffiltext{1}{National Astronomical Observatory of China, Beijing, China 100012}
\altaffiltext{2}{Harvard-Smithsonian center for Astrophysics, 60 Garden st. Cambridge, MA 02138}
\altaffiltext{3}{Eureka Scientific Inc, Oakland, CA, 94602}
\altaffiltext{4}{Nanjing University, Nanjing, China}

\begin{abstract}

The explosion of a Type Ia supernova, SN~2011ef, in the nearby Pinwheel galaxy
(M101 at $6.4$~Mpc) provides an opportunity to study pre-explosion images and
search for the progenitor, which should consist of a white dwarf (WD), possibly
surrounded by an accretion disk, in orbit with another star.  
We report on our use of deep {\sl Chandra} observations to limit the luminosity
and temperature of the pre-explosion white dwarf (WD). 
It is found that if the spectrum was a blackbody, then WDs of the highest
possible temperatures and luminosities are excluded but, even if the WD was
emitting at the Eddington luminosity, values of $kT$ less than roughly $60$~eV
are permitted.  This allows the progenitor to be an accreting nuclear-burning
WD with an expanded photosphere. 
Pre-SN HST observations were used to derive a lower limit of about $10$~eV for
the expanded photosphere. 
Li et al.\, (2011) have already ruled out the possibility of a giant donor.  We
consider the combined emission from the WD, disk, and donor, and find that even
the combined emission from a bright subgiant, WD and disk would not likely have
been observed prior to explosion,  and neither would some local candidates for
the nuclear-burning WD model. 

\end{abstract}


\section{Introduction}

Type Ia supernovae (SNe~Ia) can serve as standardizable candles to large
cosmological distances, and have enabled us to discover and study the
acceleration of Universal expansion (Riess et al.  1998; Perlmutter et al.
1999).
While there is a general consensus that an SN~Ia results  from the
thermonuclear explosion of a carbon-oxygen white dwarf (WD) with mass
approximately equal to the Chandrasekhar mass $M_{Ch}\sim1.35M_\odot$, and that
the WD must have a companion that donated mass to it, the exact nature(s) of
the companion and the mode(s) of mass transfer are still unknown (see
Hillebrandt \& Niemeyer 2000 for reviews).
In single degenerate (SD) models, the WD accretes matter from its companion.
The companion may be a main-sequence star, subgiant, or giant; it may fill its
Roche lobe and/or donate mass through winds.  
In double degenerate (DD) models, matter is donated by another WD, and there
may be a merger (e.g., Webblink 1984; Iben \& Tutukov 1984).

The explosion of an SNe~Ia (SN2011fe; Nugent et al. 2011) in M101, a
well-studied galaxy located only $6.4$~Mpc away (Shappee \& Stanek 2011) allows
us to search for the progenitor in pre-explosion images.  
Of special interest are {\sl Chandra} observations from the Megasecond survey
of M101 in 2003 (PI: Kuntz).  
SN2010fe was not detected in the {\sl Chandra} images (Liu 2008, 2011), and our
preliminary analysis placed an upper limit on X-ray luminosity of
$3-5\times10^{35}$~erg~s$^{-1}$ (Soderberg et al. 2011).  This upper limit is
two orders of magnitude lower than for previous limits on other SNe Ia.  
In this paper we use the X-ray data to constrain the bolometric luminosity of
the WD that was about to explode. 
We compute the expectations from each of several possible models, to interpret
the significance of the lack of a detectable pre-SN X-ray source.

We also predict the level of optical emission that might be expected from an
actively accreting WD and the disk that channels mass to it.
This allows us to further test the accretion model by comparing the
expectations with the results of pre-SN optical observations. 
Li et al. (2011) have used historical imaging to constrain the optical
luminosity of the progenitor to be 10-100 times fainter than previous limits on
other SN Ia progenitors. 
By doing so, they have ruled out luminous red giants and the vast majority of
helium stars as the donor, and found that any evolved red companion must have
been born with mass $\le3.5M_\odot$. 
These observations favor a model where the pre-explosion WD of SN2011fe accreted
matter either from another WD or by Roche lobe overflow from a subgiant or
main-sequence companion star.
We consider whether the optical observations place additional limits on the
accretion model.

In \S 2 we outline the expectations based on both theoretical work and the
results of prior observing programs.  In \S 3, we use a blackbody with large
ranges of bolometric luminosities and temperatures to represent the pre-SN WD,
and check whether the pre-SN {\sl Chandra} observations would have detected it.
Additional constraints from pre-SN {\sl HST} observations are also employed to
constrain the expanded photosphere and the combination of donor and accretion
disk.  We discuss the implications of our results in \S 4.

\section{Expectations}
 
Theory predicts that WDs accreting in the range 
$\sim3-6\times10^{-7} M_\odot$~yr$^{-1}$
 will burn the infalling mass
in a more-or-less steady way, 
producing luminosities of $1-2 \times 10^{38}$~erg~s$^{-1}$.
If the radius of the photosphere is not too much larger than the
radius of the WD, $k\, T$ is expected to be $70-80$~eV or
slightly higher  
(Nomoto et al. 1982; Iben 1982). 
Such sources can appear as luminous supersoft X-ray sources (SSSs; van den
Heuvel et al. 1992; Rappaport et al. 1994).
Di~Stefano \& Rappaport (1994) found that nuclear-burning WDs with these
properties could be detected in external galaxies, unless there was an
extraordinary amount of absorption. 
The amount of absorption expected in most locations in the face-on spiral
galaxy M101, for example, would not obscure the X-ray emission from
nuclear-burning near-$M_{Ch}$ WDs, as long as the photosphere is comparable in
size to the WD itself.

The possibility of detecting SNe~Ia progenitors prior to explosion inspired
searches with {\sl Chandra} and {\sl XMM-Newton} for bright, hot SSSs in
external galaxies.  
An algorithm was developed and applied to about half a dozen early-type and
late-type galaxies (Di~Stefano et al. 2003; Kong \& Di~Stefano 2003).
The results were that the numbers of SSSs in external galaxies of all types
were at least one to two orders of magnitude smaller than required to support
the hypothesis that the majority of SNe~Ia derive from nuclear-burning WDs that
are bright at X-ray wavelengths during the time they are accreting and burning
matter (Di~Stefano et al. 2006; Di~Stefano et al. 2007; Di~Stefano et al. 2009;
Di~Stefano et al.  2010ab).
A study of diffuse emission in several early-type environments also found
similar results (Gilfanov \& Bogdan 2010).

The lack of SSSs is not evidence for a lack of nuclear-burning WDs. 
A modest change in photospheric radius would shift the spectrum to longer
wavelengths, making the sources undetectable in X-ray.  In fact, there is
evidence that the local population of nuclear-burning WDs can switch ``off''
and ``on'' as SSSs when the photosphere expand or contract, even though their
bolometric luminosity remains high (e.g., CAL 83, Greiner \& Di Stefano 2002).
%
%
A more direct test is to  look for pre-SN X-ray emission from the locations of
recent SNe~Ia in nearby galaxies.  There have been no reliable detections of
pre-explosion X-rays from the site of an SN~Ia.  Upper limits  for the X-ray
luminosities have been dervived, most above a few $\times10^{38}$ erg/s. 
Given these high upper limits, it is not possible to rule out pre-SN emission
from a nuclear-burning WD (Nielsen et al. 2011).

\section{Pre-SN {\sl Chandra} and {\sl HST} Observations}


The location of SN2011fe was observed by {\sl Chandra} ten times during 2003
(PI: Kuntz) (Table~1).  The data were analyzed using uniform procedures in
our Chandra/ACIS survey of nearby galaxies (Liu 2008,2011); SN2011fe was not
detected in any of these observations.  For this paper we merged the ten
observations employing standard procedures.  The location of SN2011fe has an
equivalent total effective exposure time of 226 kiloseconds after vignetting
corrections for individual observations (Table~1).
Using CIAO 3.3, we ran {\it wavdetect} on the merged observation in the 0.3-8
keV band with scales of 1", 2", 4" and 8".  No point source is detected at the
location of SN2011fe, and no apparent feature is found by visual inspection. 

The upper limit for SN2011fe is estimated by summing up photons from ten
observations as listed in Table~1.  Photon events are extracted from the
SN2011fe location with three different circular apertures, which are the PSF
enclosing 95\% of the source photons at 0.5keV, 3.5" as the exposure weighted
PSF size, and 5" as the PSF size for large off-axis angles.
The background is carefully estimated from visually inspected surrounding
annuli or nearby circular regions if the location is close to the chip edges.
After background subtraction as listed in Table~1, we obtain a net count of
$-2.2\pm2.6$, $-0.4\pm2.6$, and $-3.1\pm3.5$ in the 0.3-8 keV for the three
apertures respectively, or $<2.2$ regardless the aperture chosen.
Here the errors are computed from the number of photon counts $N$ in the
aperture as $\sigma = \sqrt{N}$.
Because a progenitor could have a supersoft X-ray spectrum, we also estimate
the photon counts in the 0.1-8 keV band. 
There are only a few photons below 0.3 keV at the location of SN2011fe, and we
obtain a net count rate of $-3.0\pm2.8$, $0.2\pm2.6$, and $-3.9\pm3.7$ for the
three apertures respectively in the 0.1-8 keV band, or $<2.8$ regardless the
aperture chosen.

To be conservative, we place an upper limit of 3 net counts for SN2011fe.
Although the lack of photons below 0.3 keV may place stringent constraints on
the supersoft X-ray spectrum of the progenitor, we choose to adopt the 0.3-8
keV band in the following because the instrument response below 0.3 keV is
little known.
This corresponds to a count rate of $\le1.3\times10^{-5}$ c/s, or a flux of
$<5.7\times10^{-17}$ erg/s/cm$^2$ (0.3-8 keV) assuming a $kT=300eV$ blackbody
spectrum and a foreground absorption of $n_H = 1.2\times10^{20}$ cm$^{-2}$
(Dickey \& Lockman), or a luminosity of $<2.8\times10^{35}$ erg/s (0.3-8keV)
given the distance of 6.4 Mpc (Shappee \& Stanek 2011).
Assuming a lower blackbody temperature of $kT=100$ eV, this upper limit
corresponds to a luminosity of $<4.6\times10^{35}$ erg/s (0.3-8 keV).
Assuming a power-law spectrum with $\Gamma=1.7$ typical for X-ray binaries will
lead to a luminosity of $<4.9\times10^{35}$ erg/s in the 0.3-8 keV band.

The energy released through acretion and nuclear burning produce the bolometric
luminosity. Emission in X-ray wavelengths would represent only a small fraction
of the total, especially for SSSs.   
To link the count rate directly to bolometric luminosity and determine
accurately whether {\sl Chandra} observations can constrain the pre-explosion WD of
SN2011fe, we fold a blackbody spectrum through the Chandra/ACIS-S response matrix
to estimate the net counts expected in 226 kilosecond using {\tt fakeit} under
the X-ray fitting package {\tt xspec} 12.7.0e.
Figure~1 shows the expected net counts for different blackbody temperatures
and bolometric luminosities of $L_{bol} = 2\times10^{37}$, $2\times10^{38}$
and $2\times10^{39}$ erg/s.
The bolometric luminosity of $2\times10^{38}$~erg~s$^{-1}$ corresponds to the
Eddington luminosity for a $M=M_{Ch}$ WD.
The effective temperature for such a WD is expected to be 80-150 eV, and 32-600
net photons in 0.3-8 keV are expected.
Given the photon count errors of $\sigma\le3$ and upper limit of $\le3$ net
counts for SN2011fe, such a hot WD can be excluded at the $\ge10\sigma$ level.

In addition to the exclusion of a very hot WD, an important message of Figure~1
is the strong temperature dependence of the {\sl Chandra} count rates. For 
$kT\le60$~eV, the count rate is so small that even a WD emitting at
the Eddington luminosity may not have been detected in M101.  
We therefore compute the blackbody temperature above which the WD would have
been detected by pre-SN {\sl Chandra} observations with bolometric luminosities
ranging from $10^{36}$ to a few $\times10^{39}$ erg/s over a wide range of
temperatures.
This is plotted as a red solid curve for a detection threshold of 3 net counts
in the upper part of Figure~2.  
This curve will move up slightly when adopting $n_H=10^{21}$ cm$^{-2}$.
Systems in the shaded region above this curve would have been detected by
Chandra, and are thus ruled out by the non-detection.  
Systems in the unshaded region below this curve would not have been detected,
and remain viable models.  
These include WDs burning matter at the Eddington
rate, but emitting the radiation from an expanded photosphere with temperatures
lower than 60 eV.



An expanded photosphere pushes the radiation to longer wavelengths and can
therefore be constrained in the optical with pre-SN {\sl HST} observations.
As reported by Li et al. (2011), 
only upper limits can be derived for the progenitor.
To obtain these limits, we photometer the {\sl HST} observations, taken with
ACS/WFC in three filters (F435W, F555W, F814W; PI: Kuntz), using the
PSF-fitting package {\tt dolphot} (Dolphin 2000).
Based on the detected $2\sigma$ point-like objects on the same images, we
derive the $2\sigma$ detection limits in STMAG as 28.3 mag, 28.0 mag and 27.5
mag in the F435W, F555W and F814W filters, respectively.
Note that these limits are roughly 0.5 mag deeper than those derived by Li et
al.  (2011) because we use the PSF-fitting photometry in this dense stellar
field, while Li et al. make very conservative estimates with the local
background light.

To check whether the expanded photosphere would have been detected by pre-SN
{\sl HST} observations, its expected optical light is compared to the $2\sigma$
detection limits for {\sl HST} observations as shown on the $\nu F_\nu$ plot in
Figure~3.
Based on this comparison, we compute the blackbody temperature below which the
expanded photosphere would have been detected by {\sl HST} observations for
bolometric luminosities ranging from $10^{36}$ to a few $\times10^{39}$ erg/s.
This is plotted as the  blue dashed curve above the shaded region in the lower
part of Figure~2.  
Clearly, the pre-SN {\sl HST} observations would have detected an expanded
photosphere with $L_{bol} = 2\times10^{38}$ erg/s and $kT \lesssim 10 $eV, or
an expanded photosphere with $L_{bol} = 3\times10^{37}$ erg/s and $kT \lesssim
5 $eV in the shaded region, thus ruling them out.
Models in the unshaded region above the dashed curve and below the solid curve
could not be detected by either pre-SN {\sl HST} or {\sl Chandra} observations,
thus can not be ruled out.


The accretion model can be further constrained when the optical light from the
accretion disk and the donor is considered.  Li et al. (2011) have shown that
the donor could have been born with a mass no larger than 3.5$M_\odot$.
We consider a $2.5M_\odot$ subgiant, which by itself provides flux not too much
below the {\sl HST} limit.
To compute the radiation from the disk, we consider a standard multi-color disk
model but include the effects of irradiation (for a detailed discussion see
Popham \& Di~Stefano 1996).
Figure~3 shows  an example of a realistic system, including radiation from the
WD, the donor, and the disk, which cannot be ruled out.
As shown in Figure~2, the addition of the accretion disk and the donor further
excludes some expanded photosphere models that are allowed if only the
photosphere is considered.

Nature has provided some realistic systems with all the above components, with
some resembling the allowed models considered here.
As shown in Table~2 and in Figure~2, the deep Chandra observations would have
detected the M101 analog of the supersoft X-ray source CAL 87 in LMC (Greiner
et al. 1991), and the recurrent nova RS Ophi in its supersoft phase during the
outburst (e.g., Osborne et al. 2010).
On the other hand, some other SSSs, if placed in M101, could not be detected by
either Chandra or HST observations.
These include CAL 83 with a high WD mass of $1.3\pm0.3M_\odot$ (Lanz et al.
2005) and RX J0537.7-7034 (Orio et al. 1997).  Both SSSs exhibited X-ray on and
off states on time scales of months to years.

Nielsen et al. (2011) found much lower luminosity limits for the same
temperature, which would have detected even CAL 83.
However, the X-ray to bolometric luminosity conversion factors they apparently
used for spectral peak temperatures $T_{peak}=30/50/100/150$ eV were actually
the factors for the blackbody temperatures $T_{bb} = 30/50/100/150$ eV that
correspond to $T_{peak} = 2.7\times T_{bb}$.  This will lead to bolometric
luminosity upper limits $10^6/2000/16/4$ times lower.
In addition, they used exposure maps for 30/50/100/150 eV, even though the
Chandra/ACIS-S response matrix is not to be trusted or used below 0.3 keV. 
The approach we use here does not require use of the low-energy response
matrix.

\section{Discussion}


Given its relative proximity and the extensive pre-SN observations, SN2011fe in
M101 provides a unique opportunity to constrain the SNe~Ia progenitor models.
Unfortunately, though, in all SNe~Ia progenitor models the pre-SN system is
much dimmer than is typical for core collapse supernovae, making detection
difficult.  
Even in SD models, which are relatively bright, the top bolometric luminosity
of the WD is likely to be comparable to the Eddington luminosity, $\sim 2\times
10^{38}$~erg~s$^{-1}.$ 
The disk will be dimmer, typically by one or two orders of magnitude.  
Unless the donor is a giant, a possibility ruled out by Li et al.\, (2011), it
is expected to be much dimmer than the disk. 
Therefore, the existing data so far neither select a unique model nor rule out
most models.

Nevertheless, important information has been derived that places limits on the
progenitor system of SN2011fe. 
Li et al. (2011) rule out bright optical emission that would be associated with
a giant or supergiant donor.
In this paper we rule out 80~eV blackbody emission at $2\times
10^{38}$~erg~s$^{-1}$ at the $10\, \sigma$ level. 
This rules out that the progenitor WD was a Chandrasekhar mass WD accreting and
burning matter at the maximum possible rate, with a photoshpere comparable in
size to the WD's radius. 
Still allowed would be a nuclear-burning WD emitting at near Eddington
luminosity with $kT$ less than $60$~eV but higher than 10 eV, or equivalently,
with a photospheric radius 2-70 times the WD itself.
The non-detections are also consistent with WD models with lower luminosities
and slightly larger range of temperatures as shown by the unshaded region in
Figure~2.
Also allowed are some local candidates for nuclear burning WDs as listed in
Table~2.

Other models with little X-ray emission cannot be rule out by existing pre-SN
observations.
A recurrent nova in quiescence, for example,has a luminosity of 10-100$L_\odot$
(e.g., Zamanov et al. 2010), and could not be detected by pre-SN observations.
The pre-explosion WD in the DD scenario may also be dim in the epoch prior to
explosion (Dan et al. 2011). 
A fast rotating WD spun up by the accretion in either the SD or DD models may
exceed $M_{Ch}$, and will explode as SN Ia only when it spins down. This spin
down phase may take a long time without accretion from the donor, thus no X-ray
emission from hydrogen burning (e.g., Di Stefano et al. 2011).

\acknowledgements

We thank Alicia Soderberg, Robin Ciardullo, and Kim Herrmann for helpful
discussions.


\begin{deluxetable}{crcccrccccc} 
\tablecaption{Pre-SN {\sl Chandra} Observations for SN2011fe\tablenotemark{a}} 
\tablehead{ 
\colhead{OBsID} & \colhead{Texpt} & \colhead{OAA} & \colhead{PSF} & \colhead{VigF} & \colhead{Teff} & \colhead{C$_{100}$} & \colhead{C$_{300}$} & \colhead{B$_{100}$} & \colhead{B$_{300}$} & \colhead{BArea} 
} 
 
\startdata

4732.s3 & 70691 & 6.49 & 5.9 & 0.197 & 13926 & 1, 1, 1  & 1, 1, 1 & 23 & 17 & 400\\ 
4733.s3 & 25132 & 4.33 & 3.1 & 0.296 & 7439  & 0, 0, 1  & 0, 0, 1 & 23 & 21 & 400\\ 
4735.s3 & 29148 & 3.55 & 2.4 & 0.910 & 26525 & 0, 0, 0  & 0, 0, 0 & 19 & 14 & 300\\ 
5300.s3 & 52761 & 6.83 & 6.3 & 0.801 & 42262 & 4, 1, 2  & 4, 1, 2 & 30 & 28 & 300\\ 
5309.s3 & 71679 & 6.50 & 5.9 & 0.180 & 12902 & 1, 0, 1  & 1, 0, 1 & 30 & 24 & 400\\ 
5323.s3 & 43161 & 4.26 & 3.0 & 0.375 & 16185 & 1, 1, 2  & 0, 0, 1 & 12 & 12 & 400\\ 
5339.s3 & 14505 & 2.08 & 1.4 & 0.177 & 2567  & 0, 0, 1  & 0, 0, 1 & 8  & 6  & 300\\ 
6114.s3 & 67052 & 3.56 & 2.4 & 0.910 & 61017 & 1, 4, 4  & 1, 3, 3 & 75 & 63 & 300\\ 
6115.s3 & 36217 & 3.56 & 2.4 & 0.910 & 32957 & 0, 1, 1  & 0, 1, 1 & 15 & 13 & 300\\ 
6118.s3 & 11606 & 3.55 & 2.4 & 0.909 & 10550 & 0, 1, 1  & 0, 1, 1 & 9  & 7  & 300\\ 
 
\enddata 

\tablenotetext{a}{The columns are (1) Observation ID and the chip SN2011fe is
on, (2) the exposure time of the observation in seconds, (3) off-axis angle in
arcminutes, (4) the PSF size in arcseconds that encloses 95\% of the photons at
0.5 keV, (5) the vignetting factor, (6) the vignetting-corrected effective
exposure time, (7) the photon counts in the 0.1-8 keV band within separate PSFs
for the ten observations, within 3.5" circles for all observations, and within
5" for all observations, (8) the three photon counts in the 0.3-8 keV band, (9)
the background counts in the 0.1-8 keV band, (10) the background counts in the
0.3-8 keV bands, and (11) the area in pixels of the background regions used.}

\end{deluxetable}

\begin{deluxetable}{lcccccccc}
\tablecaption{Known sources with supersoft X-ray spectra\tablenotemark{a}}
\tablehead{
\colhead{Name\tablenotemark{b}} & \colhead{Type} & \colhead{Period} & \colhead{kT} & \colhead{$L_{bol}$} & \colhead{Mv} & \colhead{B-V} & \colhead{Chandra?} & \colhead{HST?}
}

\startdata

CAL 83 & CBSS & 1.04 & 39-60 & $<20$  & -1.3 & -0.06 & no  & no \\
CAL 87 & CBSS & 0.44 & 63-84 & 60-200  & 0.1 & 0.1 & yes & no \\
RX J0019.8+2156 & CBSS & 1.0-1.35 & 25-37 & 30-90 & 0.6 & 0.1 & no & no \\
RX J0513.9-6951 & CBSS & 0.76 & 30-40 & 1-60 & -2.0 & -0.1 & no & yes \\
RX J0537.7-7034 & CBSS & 0.13 & 18-70 & 6-20 & 1.0 & -0.03 & no & no \\
RS Ophi & RN & 453.6 & 65-95 & 114-950 & -5 & 0.5 &  yes & yes \\

\enddata

\tablenotetext{a}{The information for the first five close binary supersoft
sources (CBSS) are taken from Greiner et al. (1996).
For the recurrent nova (RN) RS Ophi, the X-ray
information in the brief (duty cycle: 1\%) supersoft phase is taken from
Osborne et al. (2011), the period is taken from Bandi et al. (2009), $M_V$ and
B-V during supesoft X-ray phase are taken from Hric et al. (2008).}
\tablenotetext{b}{The columns are (1) the source name, (2) the source type, (3)
the period in days, (4) the temperature in eV, (5) the bolometric luminosity in
$10^{36}$ erg/s, (6) the absolute V magnitude, (7) B-V, (8) whether it would
have been detected by pre-SN Chandra observations if placed in M101, and (9)
whether it would have been detected by pre-SN HST observations. }

\end{deluxetable}

\begin{figure}

\plotone{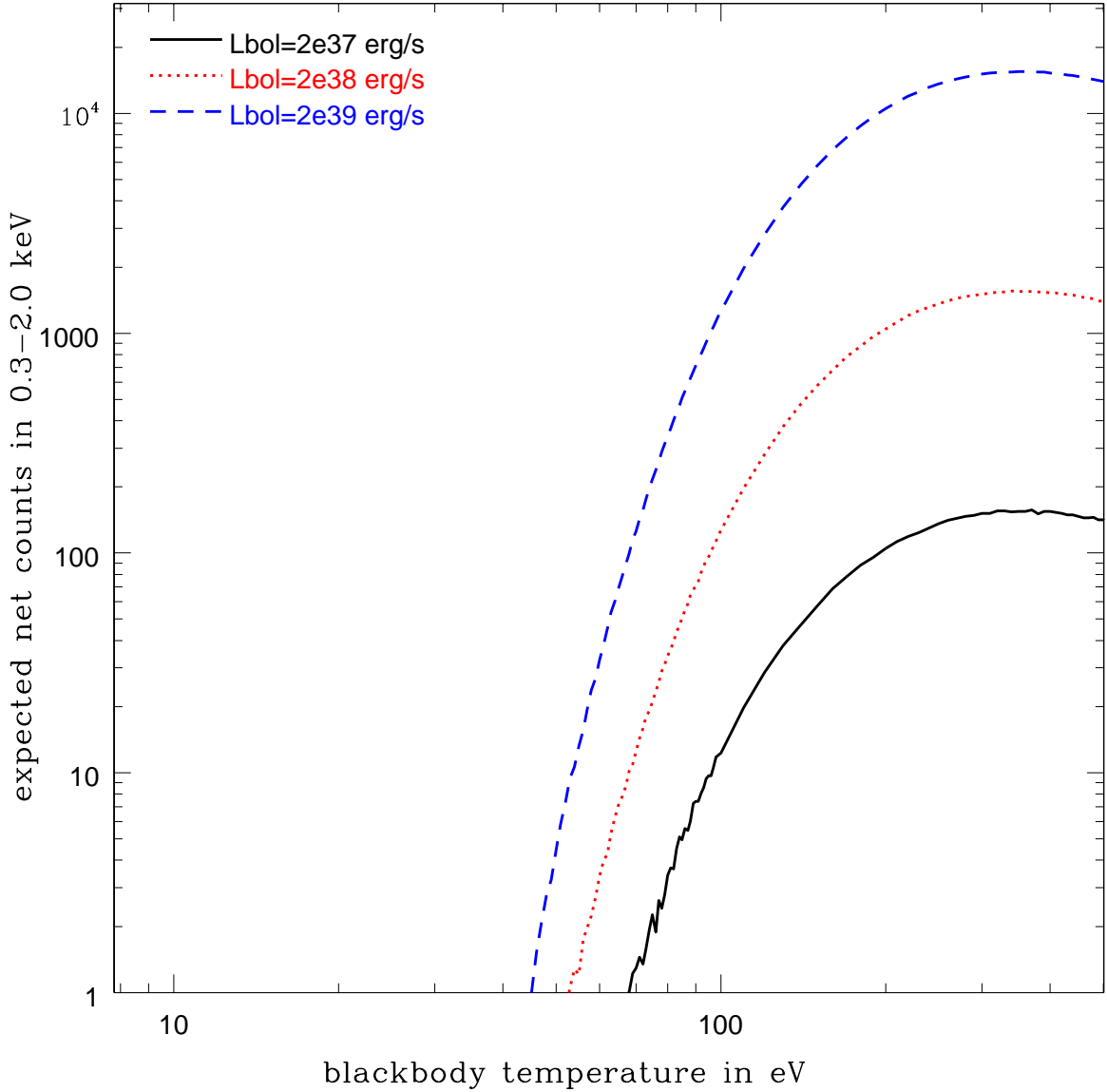}

\caption{The expected net counts in 0.3-2.0 keV for a blackbody spectrum with
different temperatures under bolometric luminosities of $L_{bol} =
2\times10^{37}$, $2\times10^{38}$ and $2\times10^{39}$ erg/s when observed by
the merged pre-SN {\sl Chandra} observations. }

\end{figure}

\begin{figure}

\plotone{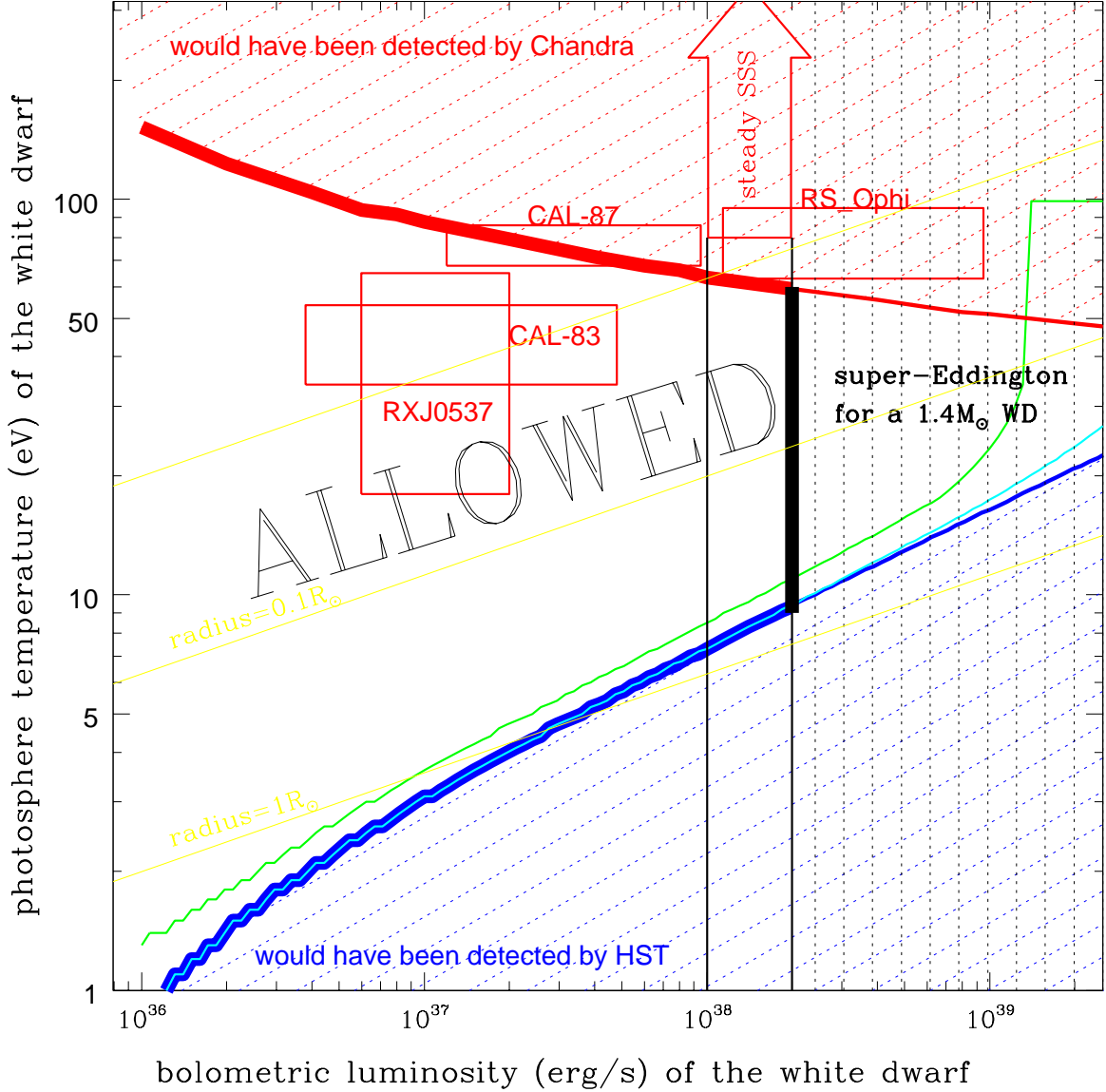}

\caption{Constraints on the temperature and the bolometric luminosity of the
pre-explosion WD by pre-SN {\sl Chandra} and {\sl HST} observations. The unshaded
region to the left enclosed by the thickened curves represents models allowed
by the non-detections in both {\sl Chandra} and {\sl HST} observations.  
The upper red thick solid curve stands for the temperature above which a
blackbody would have been detected by pre-SN {\sl Chandra} observations; the
lower blue thick solid curve stands for the temperature of an expanded
photosphere below which the photosphere would have been detected by pre-SN {\sl
HST} observations.
Similar to the blue curve are the cyan line with optical light also from the
irradiated accretion disk considered, and the green line with optical light
also from the disk and the donor (an evolved star of $2.5M_\odot, 12R_\odot$ and 630Myrs old
in this case) considered.  The red arrow indicates the location of a $M =
M_{Ch}$ WD with steady nuclear burning on its surface.  The various boxes
indicate the locations for known supersoft sources as listed in Table~2.  }

\end{figure}

\begin{figure}

\plotone{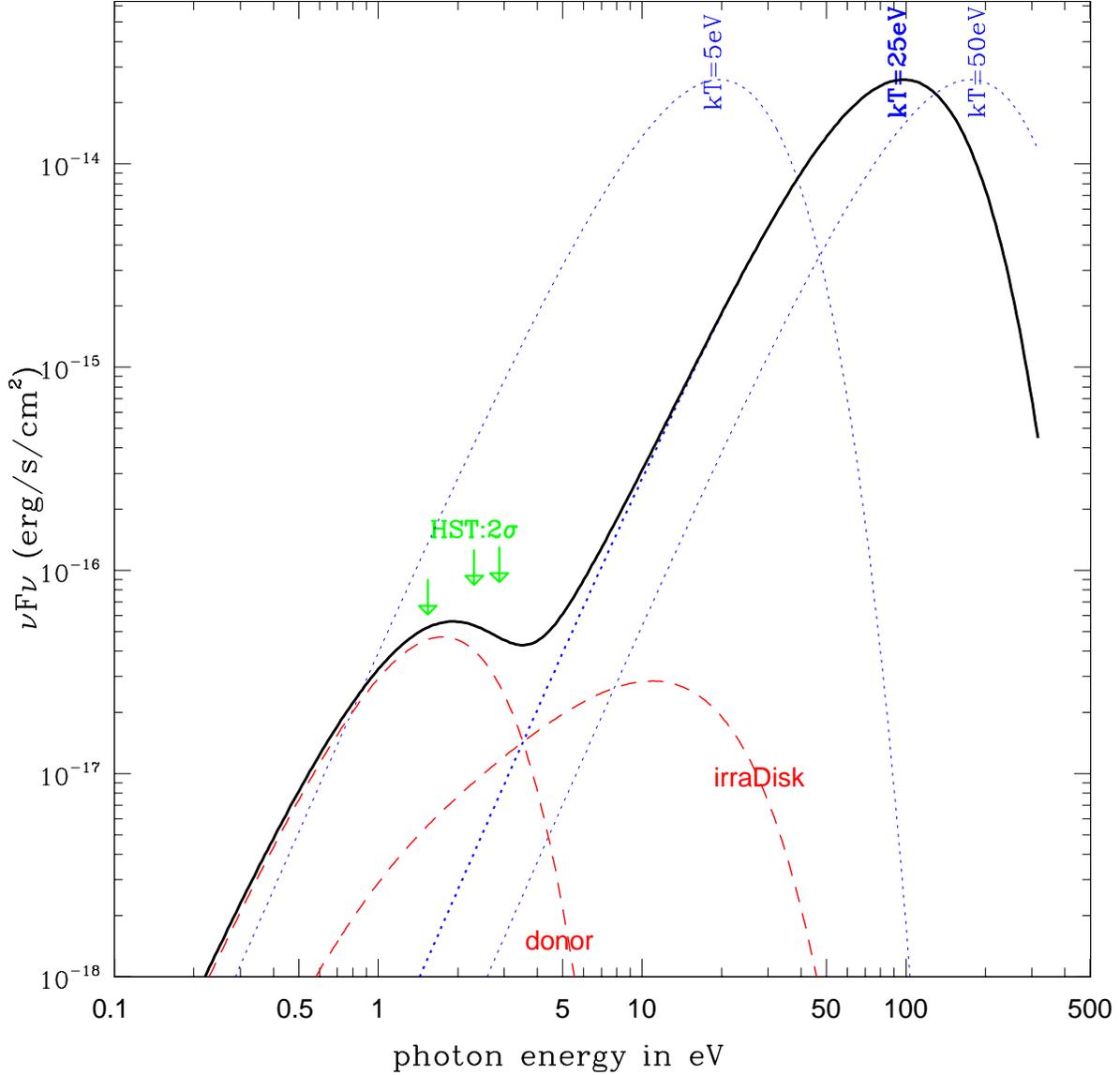}

\caption{Expected optical light from different components of a possible
SN2011fe progenitor as compared to the $2\sigma$ detection thresholds for
pre-SN {\sl HST} observations.  The dotted curves stand for the spectra for an
expanded photosphere with temperatures of 5 eV, 25 eV and 50 eV.  The dashed
curves, from left to right, stand for the spectrum for the donor (an evolved
star of $2.5M_\odot, 12R_\odot$ and 630Myrs old in this case) and the spectrum
for the irradiated accretion disk around a 25 eV photosphere.  The solid curve
stands for the summed radiation from the photosphere, the irradiated accretion
disk, and the donor.  Overplotted for comparison are known supersoft sources
observed in the optical.  }

\end{figure}

\end{document}